\documentstyle[12pt]{article}

\begin{document}
\bibliographystyle{unsrt}

\topmargin 0pt
\oddsidemargin 5mm
\renewcommand{\thefootnote}{\fnsymbol{footnote}}

\begin{titlepage}
\setcounter{page}{0}
\rightline{Preprint YERPHY--1446(16)--95}

\vspace{2cm}
\begin{center}
{\Large Pseudoclassical theory of Mayorana--Weyl particle}
\vspace{1cm}

{\large Grigoryan G.V.\raisebox{.8ex}{$\star$}, Grigoryan R.P.
\raisebox{.8ex}{$\star\star$}, Tyutin
I.V.\raisebox{.8ex}{\ddag}
\footnote{ P.N.Lebedev Physical Institute, Moscow, Russia}}
\footnote{ Partially supported  by the grant 211-5291 YPI
 of the German Bundesministerium f\"ur Forschung und
 Technologie.}\\
\vspace{1cm}

{\em Yerevan Physics Institute, Republic of Armenia}\\

\vspace{1cm}
\raisebox{.8ex}{$\star$}{E-mail:GAGRI@VXC.YERPHY.AM}\\
\raisebox{.8ex}{$\star\star$}{E-mail:ROGRI@VXC.YERPHY.AM}\\
\raisebox{.8ex}{\ddag}\vspace{1cm}
{E-mail:TYUTIN@LPI.AC.RU}
\vspace{5mm}

\end{center}
\centerline{{\bf{Abstract}}}
A pseudoclassical theory of Weyl particle in the space--time
dimensions $D=2n$ is constructed. The canonical quantization of
that pseudoclassical theory is carried out and it results in the
theory of the $D=2n$ dimensional Weyl particle in the
Foldy--Wouthuysen representation. A quantum mechanics of the
neutral Weyl particle in even--dimensional space--time is suggested
and the connection of this theory  with the theory of
Mayorana--Weyl particle in QFT is discussed for $D=10$.

\vfill
\centerline{\large Yerevan Physics Institute}
\centerline{\large Yerevan 1995}

\end{titlepage}
\newpage
\renewcommand{\thefootnote}{\arabic{footnote}}
\setcounter{footnote}{0}

\section{Introduction}
\indent
In spite of the bulk of the papers devoted to the theories of point
particles and to methods of their quantization, the problem still
attracts the attention of the investigators and numerous
classical models of particles and superparticles were discussed
recently. The renewal of the interest to these theories is
primarily due to the problems in the string theory since particles
are prototypes of the strings.

As it is known, the theory of RNS string with a GSO projection
(see. \cite{GSW}) is a supersymmetric theory in ten dimensional
space--time. The supersymmetry requires that each mass level
comprises a supermultiplete. In the massless sector the
superpartners to the gauge vector fields , which in $D=10$
dimensions have eight degrees of freedom, are
massless fermions -- Mayorana--Weyl bispinors.  Hence it is
interesting to construct the classical (pseudoclassical) model,
which after quantization will bring to the theory of the
Mayorana--Weyl bispinor in $D=10$ dimensions.

The first pseudoclassical description of the relativistic
spinning particle was given in paper \cite{BM1,BM2}. It was
followed by a great number of papers  [4-23]
 devoted to the different
quantization schemes of that theory, to the introduction of the
internal symmetries and to the generalization to higher spins.

A pseudoclassical theory of Weyl particle in the space--time
$D=4$ was constructed in\cite{GGT}. In this paper the method
suggested in \cite{GGT} is generalized  to the arbitrary even
dimensions $D=2n$ .This is realized in sect.2, where
the canonical quantization of
that pseudoclassical theory is carried out and it results in the
theory of the $D=2n$ dimensional Weyl particle in the
Foldy--Wouthuysen representation. A quantum mechanics of the
 neutral Weyl particle in even--dimensional space--time is presented
in sect.3
and the relation of this theory  to the theory of
Mayorana--Weyl bispinor in QFT is discussed in sect.4 in $D=10$
dimensions.

\section{$D=2n$--dimensional Weyl particle}
\indent

Consider a theory with the action given by the expression
\begin{eqnarray}
\label{ACT}
S=\int d\tau\left[\frac{1}{2e}\left(\dot{x}^\mu -
\frac{i}{2}\chi\xi^\mu -
\frac{(-i)^{\frac{D-2}{2}}}{(D-2)!}
\varepsilon^{\mu\nu\lambda_1\dots \lambda_{D-2} }
b_\nu\xi_{\lambda_1}\cdots \xi_{\lambda_{D-2}}+\tilde{\alpha} b^\mu
\right)^2-
\frac{i}{2}\xi_\mu\dot{\xi}^\mu \right],
\end{eqnarray}
which is a generalization to the space-time dimension $D=2n$
of  the  pseudoclassical theory of Weyl particle \cite{GGT}.
Here   $x_\mu$   are   the   coordinates  of  the  particle,
$\mu=0,1,\dots  ,D-1$;  $\xi^\mu$  are  Grassmann  variables
describing  the  spin degrees of freedom; $e,\,\chi,\,b^\mu$
are  additional  fields,  $\tilde{\alpha}$ is a constant; $e,\,b^\mu
,\,\tilde{\alpha}$  being  grassmann  even,  $\chi$-  grassmann  odd
variables;   the  overdote  denotes  a  differentiation  with
respect to the parameter $\tau $ along the trajectory.

The  action  (\ref{ACT})  is  invariant under three types of
gauge   transformations:  the  re\-pa\-ra\-me\-tri\-zation
transformations
\begin{eqnarray}
\label{RT}
\delta x^\mu=u\dot{x}^\mu,  \quad\delta
e&=&\frac{d}{d\tau}(ue),  \quad\delta b^\mu=
\frac{d}{d\tau}(ub^\mu),\nonumber\\
\delta\xi^\mu&=&u\dot{\xi}^\mu,  \quad
\delta\chi=\frac{d}{d\tau}(u\chi)
\end{eqnarray}
with    the    even    parameter    $u(\tau)$,    supergauge
transformations
\begin{eqnarray}
\label{ST}
\delta x^\mu&=&iv\xi^\mu,\quad\,\delta e=iv\chi,\quad\,\delta
b^\mu=0,\,\quad
\delta\xi^\mu=v\frac{z^\mu}{e},\quad\,\delta \chi=2 \dot{v},\nonumber\\
z^\mu&=& \dot{x}^\mu -
\frac{i}{2}\chi\xi^\mu -
\frac{(-i)^{\frac{D-2}{2}}}{(D-2)!}
\varepsilon^{\mu\nu\lambda_1\dots \lambda_{D-2} }
b_\nu\xi_{\lambda_1}\cdots \xi_{\lambda_{D-2}}+\tilde{\alpha} b^\mu
\end{eqnarray}
with   the   odd   parameter   $v(\tau)$,   and  also  under
transformations \cite{GGT}
\begin{eqnarray}
\label{WT}
\delta x^\mu&=&\frac{(-i)^{\frac{D-2}{2}}}{(D-2)!}
\varepsilon^{\mu\nu\lambda_1\dots \lambda_{D-2} }\eta_\nu
\xi_{\lambda_1}\cdots \xi_{\lambda_{D-2}}-\tilde{\alpha}
\eta^\mu,\nonumber\\
\delta \xi^\mu&=&\frac{1}{e}\frac{(-i)^{\frac{D}{2}}}{(D-3)!}
\varepsilon^{\mu\nu\delta\lambda_2\dots \lambda_{D-2} }
\eta_\nu z_\delta\xi_{\lambda_2}\cdots
\xi_{\lambda_{D-2}},\\
\delta b^\mu&=&\frac{d}{d\tau}(\eta^\mu),
\quad\delta\chi=-2\eta_\nu(p^\nu
\xi^\sigma-p^\sigma \xi^\nu)b_\sigma \delta_{D4},  \quad \delta e=
-2i\eta_\nu \xi^\nu\xi^\sigma b_\sigma \delta_{D4},\nonumber
\end{eqnarray}
with the even parameter $\eta_\nu(\tau)$.

Acting in the standard way \cite{D1, GTY} we obtain the canonical
hamiltonian of the theory, which is given by the expression
\begin{eqnarray}
\label{CH}
H &=&\dot{x}^\mu p_\mu+\dot{\xi}^\mu\pi_\mu
-L=\nonumber\\
&=&\frac{e}{2}p^2+\frac{i}{2}\chi p_\mu \xi^\mu-
\left(\frac{(-i)^{\frac{D-2}{2}}}{(D-2)!}
\varepsilon^{\nu\mu\lambda_1\dots \lambda_{D-2} }
p_\mu\xi_{\lambda_1}\cdots \xi_{\lambda_{D-2}}+\tilde{\alpha}
p^\nu\right)b_\nu,
\end{eqnarray}
primary constraints
\begin{equation}
\label{CNST}
\Phi_1^{(1)}=\pi_e, \quad \,  \Phi_2^{(1)}=\pi_\chi, \,\quad
\Phi_{3\mu}^{(1)}=\pi_\mu-\frac{i}{2}\xi_\mu,
\quad\,\Phi_{4\mu}^{(1)}=\pi_\mu^b
\end{equation}
 and  the secondary constraints
\begin{eqnarray}
\label{SC}
\Phi_1^{(2)}&=&|p_0|-\omega,  \quad
\Phi_{2}^{(2)}=p_\mu\xi^\mu,\nonumber\\
\Phi_{3\mu}^{(2)}&=&T_\mu=\frac{(-i)^{\frac{D-2}{2}}}{(D-2)!}
\varepsilon_{\mu\nu\lambda_1\dots \lambda_{D-2} }
p^\nu\xi^{\lambda_1}\cdots \xi^{\lambda_{D-2}}+\tilde{\alpha} p_\mu,
\end{eqnarray}
 where $\omega=|\vec{p}|,\,\vec{p}=(p_k),\,k=1,\dots,D-1$.
One can see now, that the canonical hamiltonian $H$ is
equal  to zero  on  the  constraints  surface, as it was
expected to be.

The  constraints $F\equiv (\Phi^{(1)}_1,  \,  \Phi^{(1)}_2,  \,
\Phi^{(1)}_{4\mu},  \,  \Phi^{(2)}_1),  \,  \Phi^{(2)}_{3\mu}$
 are  first class. Apart from them there is
one more  first class constraint $\varphi$, which is a
linear combination of the constraints
$\Phi_{(3\mu)}^{(1)}$,  $\Phi^{(2)}_2$:
\begin{equation}
\label{FCC}
\varphi=p^\mu \Phi_{3\mu}^{(1)}+i\Phi_2^{(2)}=p_\mu
\pi^\mu+\frac{i}{2}p_\mu \xi^\mu.
\end{equation}
Adhering  to  the  quantization  method, when already at the
classical  level  all  gauge  degrees  of  freedom are fixed
\cite{GT2}  ,  we  must introduce into the theory additional
constraints,  equal  in  number  to  that of all first class
constraints and conjugated to the latter. However, as it was
noted in \cite{GGT}, the constraints $\Phi_{3\mu}^{(2)}$ for
$D\geq 4$, are at least quadratic functions of the variables
$\xi^\lambda$,   thus   complicating  the  introduction  of
additional constraints conjugated to $\Phi_{3\mu}^{(2)}$. For
this  reason,  following \cite{GGT}, the constraints
$\Phi_{3\mu}^{(2)}$  after quantization will be used as
conditions on the state vectors.  For  the remaining first class
$F$ and $\varphi$ constraints we will introduce equal number of
additional constraints $\Phi^G$ in the form

\begin{equation}
\label{GF}
\Phi^G_{1}=x_0-\kappa \tau, \quad \,  \Phi^G_{2}=\chi, \quad \,
\Phi^G_{3}=e-\frac{\kappa}{p_0},\quad\, \Phi^G_{4\nu}=b_\nu, \quad \,
\Phi^G_{5}=\xi_0,
\end{equation}
where    $\kappa=-{\rm sign}p_0$.   The   constraint
$x_0-\kappa\tau\approx 0$ was introduced in \cite{GT2} as one,
conjugated to the constraint $|\vec{p}|- \omega\approx 0$;
$\kappa=+1$ corresponds to the particle sector in
the  theory,  while $\kappa=-1$ corresponds to the
antiparticle sector. Now  the  constraints  (\ref{CNST}),
(\ref{SC}) (except for $\Phi_{3\mu}^{(2)}$)  and  (\ref{GF})
constitute a system of second class constraints. To use the
Dirac method of   quantization   of   the  theories  with  second
class constraints, we'll pass to a set of constraints, which do
not depend  on  time  explicitly. To go over to time independent
set  of  constraints  we  perform a canonical transformation from
the variables $x^\mu,  \,  p_\mu$ to variables
$x^{\mu\prime},\,p^\prime_\mu$, defined by the relations
\begin{equation}
\label{PV}
x^{\prime}_0=x_0-\kappa\tau,  \quad x^{i\prime}=x^i\quad
p^\prime_\mu=p_\mu
\end{equation}
( corresponding generating function is given by {\mbox{$W
= x^\mu p^\prime_\mu-\tau\kappa p^\prime_0$}} ).
The  constraint  $\Phi_1^G$  in terms of new variables takes
the form $x_0^\prime\approx 0$; all other constraints remain
unchanged.  The  hamiltonian of the system on the constraint
surface is given by the expression
\begin{equation}
\label{HAM}
H=\omega=|\vec{p}|.
\end{equation}

     To  find  the  Dirac brackets for independent variables
$x^i,\,  p_j,  \, \xi^k$ of the theory we'll make use of the
results  of the paper \cite{GG1}, where the theory of $D=2n$
dimensional  relativistic  spinning particle was considered. Note
,  that  in  comparison with the paper \cite{GG1}, the system  of
second--class constraints in this theory contains new constraints
$\Phi_{4\mu}^{(1)}=\pi_\mu^b\approx  0,$ $\,\Phi^G_{4\mu}=
b_\mu\approx   0$.   However  they  have  a special  form
\cite{GTY} and  they  do  not  affect  the  final Dirac brackets
(the  variables  $b_\mu$  and  $\pi_\mu^b$  can be excluded from
the theory using the constraints). This allows to  use
immediately the results of \cite{GG1}, taken in the massless
limit, to obtain the following final Dirac brackets

\begin{eqnarray}
\label{DB}
\{x^i,  x^j\}_D&=&\frac{i}{2\omega^2}[\xi^i,\xi^j]_-,  \quad
\{x^i,  p_j\}_D=\delta^i_j,  \quad\{p_i,  p_j\}_D=0,  \nonumber\\
\{x^i,  \xi^j\}_D&=&\frac{1}{\omega^2}\xi^i p^j ,\quad
\{\xi^i,  \xi^j\}_D=-i\left(\delta^{ij}-\frac{p^i
p^j}{\omega^2}\right),  \quad
\{p_i,  \xi^j\}_D=0.
\end{eqnarray}
The quantization of the theory is carried out through the
realization of the operators $x^i,\,\hat{p}_i,\,\xi^i$ in the
form \cite{GT2, GG1}:
\begin{eqnarray}
\label{QOR}
\hat{x}{}^i&=&\hat{q}{}^i-\frac{i\hbar}{4
\hat{\omega}^2}\left[\Sigma^i,\hat{p}_k
\Sigma^k\right]_-,\quad\nonumber\\
\hat{\xi}{}^i&=&\left(\frac{\hbar}{2}\right)^{1/2}\hat{\kappa}
\left[\Sigma^i-\frac{1}
{\hat{\omega}^2}\hat{p}_i\hat{p}_j
\Sigma^j\right],\quad \hat{p}_k= -i\frac{\partial}{\partial q^k},
\end{eqnarray}
where the operators $\hat{q}^i$ (physical coordinate operators)
are multiplication operators, the variable $\kappa$ is replaced
by the operator $\hat{\kappa}$
  \begin{equation}
\label{KAP}
\hat{\kappa}=\left(
\begin{array}{cc}
I&0\\0&-I
\end{array}
\right)=\gamma^0=\tau^3\otimes I, \quad \hat{\kappa}^2=1
\end{equation}
with eigenvalues $\kappa=\pm 1$, the operators $\Sigma^i={\rm
diag (\sigma^i, \sigma^i)}$, $\sigma^i$ are $2^{(D-2)/2}\times
2^{(D-2)/2}$ matrices, which realize the Clifford algebra
$\left[\sigma^i,\sigma^j\right]_+=\delta^{ij}I$.

     Now    using    the    expressions    (\ref{QOR})   for
$\hat{\xi}^i$,  we can find the expressions for the operators
$\hat{T}_\mu$,   which   correspond   to   the  first  class
constraints $\Phi_{3\mu}^{(2)}$:
\begin{equation}
\label{TO}
\hat{T}_{\mu}=\left(\frac{\hbar}{2}
\right)^{(D-2)/2}\hat{p}_\mu \hat{T},\quad
\hat{T}=\hat{\kappa}\frac{\hat{p}^i \Sigma^i}{\hat{\omega}}
-\alpha,  \quad\hat{p}_0=-\hat{\kappa}\hat{\omega},
\quad \alpha=\left(\frac{\hbar}{2}\right)^{-(D-2)/2}\tilde{\alpha},
\end{equation}
To deduce these relations we used the equality
$\varepsilon_{01\dots(D-1)}=-\varepsilon_{12\dots(D-1)}=-1$,
and also the relation
\begin{equation}
\frac{(-i)^{\frac{D-2}{2}}}{(D-2)!}
\varepsilon^i_{\,j_1\cdots j_{D-2} }
\sigma^{j_1}\cdots \sigma^{j_{D-2}}=\sigma^i
\end{equation}
from the $\sigma^i$--matrix algebra in $(D-1)$--dimensional space
\cite{CA}.

The canonical generators of the Lorentz transformation
\begin{equation}
\label{CLG}
J^{\mu\nu}=-\left(x^\mu p^\nu-x^\nu p^\mu
+\frac{i}{2}[\xi^\mu,\xi^\nu]_-\right)
\end{equation}
after  quantization  of the theory, in terms of operators of
the physical variables, are given by the expressions \cite{GG1}
\begin{eqnarray}
\label{LG}
\hat{J}^{ik}&=&-\hat{q}^i \hat{p}^k + \hat{q}^k \hat{p}^i -
\frac{i\hbar}{4}\left[\Sigma^i, \Sigma^k\right]_{-},\nonumber\\
\hat{J}^{0k}&=&-x^0 \hat{p}^k
-\frac{1}{2}\hat{\kappa}\left[\hat{q}^k, \hat{\omega}\right]_{+}
-\frac{i\hbar}{4 \hat{\omega}}\hat{\kappa}\hat{p}^j
\left[\Sigma^k, \Sigma^j\right]_- .
\end{eqnarray}
It's not difficult to check that the $\hat{J}^{\mu\nu}$ operators
commute with the operator $\hat{T}$. Furthermore, introducing the
projection operators
\begin{equation}
\label{TP}
\hat{T}_{\pm}=-\frac{1}{2}\alpha
\hat{T}|_{\alpha=\pm}=\frac{1}{2}\left(1\mp
\hat{\kappa}\frac{\hat{p}^i\Sigma^i}{\hat{\omega}}\right),
\end{equation}
which commute with $\hat{J}^{\mu\nu}$, we can represent the
$\hat{J}^{\mu\nu}$ operator in the form
\begin{equation}
\label{JT}
\hat{J}^{\mu\nu}=\hat{T}_+\hat{J}^{\mu\nu}
+\hat{T}_-\hat{J}^{\mu\nu} =
\hat{T}_+\hat{J}^{\mu\nu}\hat{T}_+
+\hat{T}_-\hat{J}^{\mu\nu}\hat{T}_- ,
\end{equation}
 which  reflects the property of the reducibility of the
 representation  of the Lorentz group for $m=0$.

As   it   was   already   mentioned   above,  the  operators
$\hat{T}_\mu$   will  be  used to impose conditions on the
physical  state  vectors. For the $D$--dimensional space--time the
state  vector  in  general  has  $2^{D/2}$ components .
Representing the state vector $f$ in the form

 \begin{equation}
\label{SV}
f=\left(
\begin{array}{c}
f^1(q)\\f^2(q)
\end{array}
\right),\quad q = (x^0,q^i),
\end{equation}
where    $f^1(q)$   and   $f^2(q)$   are   $2^{(D-2)/2}$
component spinors, we  write  down  the  equations  for
 the state vector in the  form:
\begin{equation}
\label{TO2}
\hat{T}f=0.
\end{equation}

It is natural to interpret the quantum mechanics constructed
above as a theory of Weyl particle in the Foldy--Wouthuysen
representation. Indeed, consider  the Schr\"odinger equation
$(i\partial/\partial\tau-\hat{H})  f=0$, which describes the
evolution  of the state-vector $f$ with respect to parameter
$\tau$ . Being rewritten in terms of the physical time
$x_0=\kappa\tau$  it takes the form
\begin{equation}
\label{SE}
(i\frac{\partial}{\partial x^0}-\gamma^0 \hat{\omega})f=0.
\end{equation}

Applying  the  unitary  Foldy--Wouthuysen  transformation for
the case of massless particle in  $D$  dimensional  space--time
\begin{equation}
\label{FW}
 f=U \psi,  \quad
U=\frac{\hat{\omega}+\gamma^i\hat{p}^i}{\hat{\omega}\sqrt{2}},
\end{equation}
 where $\psi$ is the wave function in the Dirac representation,
 $\gamma^\mu=(\gamma^0,\gamma^i)$ are Dirac $\gamma$-- matrices
 in $D=2n$--dimensional space--time,
 we find \cite{GT2,GG1} that the  Schr\"odinger equation
transforms into Dirac equation, the expressions for the Lorentz
generators $\hat{J}^{\mu\nu}$ transform into standard expressions
for the Lorentz generators in the Dirac representation.
Furthermore, by direct calculation one can prove that the
operator
\begin{equation}
\label{TDO1}
\hat{T}\equiv\hat{T}_{FW}=\gamma^0\frac{\hat{p}^i \Sigma^i}
{\hat{\omega}}-\alpha
\end{equation}
 transforms into the $\hat{T}_D$ operator
\begin{equation}
\label{TDO}
\hat{T}_D=U^+\hat{T}_{FW} U=\gamma^{D+1} - \alpha,
\end{equation}
where $\gamma^{D+1}$ is the analogue of Dirac $\gamma^5$--matrix
in dimensions $D=2n$. One can see, that operator $\hat{T}_D$ is
proportional to a standard Weyl projector in the Dirac
representation.

Thus we see that the quantum mechanical description constructed
here after the Foldy--Wouthuysen transformation  turns into the
Dirac description of the Weyl particle. Hence the above
constructed quantum mechanics  describes the Weyl particle in the
Foldy--Wouthuysen representation.

\section{ Quantum mechanics of Mayorana--Weyl spinor}
\indent

     Now  we  will  proceed  with  the  construction  of the
pseudoclassical  theory  of  Mayorana--Weyl particle.
  Note, that the action (\ref{ACT}) is invariant
under the transformations
\begin{eqnarray}
\label{TI}
x^\mu(\tau)&\rightarrow& x^\mu(-\tau),   \quad
\xi^\mu(\tau)\rightarrow \xi^\mu(-\tau),   \quad
\xi_{D+1}(\tau)\rightarrow
-\xi_{D+1}(-\tau),\nonumber\\
\chi(\tau)&\rightarrow&\chi(-\tau),\quad  e(\tau)\rightarrow
e(-\tau),\quad b^\mu(\tau)\rightarrow -b^\mu(-\tau),\quad
i\rightarrow -i,
\end{eqnarray}
which correspond to the reparametrization $\tau\rightarrow
-\tau$. This transformation was not included in the gauge
group in the previous section. In that case
the model describes the charged particle and in the gauge
$x^0-\kappa\tau\approx 0$ the trajectories with $\kappa=+1$ are
interpreted  as  trajectories  of  particles  and those with
$\kappa=-\tau$ as trajectories of
antiparticles.  The switching  on of the external electromagnetic
field confirms this assertion since  to the trajectory with a
given $\kappa$ corresponds a particle  with a  charge  $\kappa
e$  \cite{GT2} and the action isn't invariant under the
transformation $\tau\rightarrow -\tau$. When the action is
invariant under the transformation $\tau\rightarrow -\tau$,
there is a possibility of another interpretation. We can
identified the trajectories with $\kappa=+1$ and $\kappa=-1$.
This is equivalent to the introduction of the reparametrization
$\tau\rightarrow -\tau$ in the gauge group \cite{GT2} and then
the theory describes the truly neutral particle.

Thus to describe Mayorana--Weyl spinor we enlarge the symmetry
group   of   the  action  (\ref{ACT}), given  by  the  gauge
transformations (\ref{RT})--(\ref{WT}), by the transformation
(\ref{TI}) and  will  choose  for  $\Phi_1^G$  (see.(\ref{GF}))
the constraint $\Phi_1^{\prime G}=x_0-\tau$.

Formally  this  means  that now we must take all equations of
the  previous  section  for  the  value  of  $\kappa=+1$.
The wave functions described now by a $2^{(D-2)/2}$
column $f=f^1$. The hamiltonian is equal to $H=\hat{\omega}$
and the realization of the operators $\hat{x}^i,\,\hat{p}_i,\,
 \hat{\xi}^i$ are given by (\ref{QOR}) with $\kappa=+1$.

Now with $\kappa=+1$ the expression for operators $\hat{T}$ and
$\hat{J}^{\mu\nu}$  take the form
\begin{equation}
\label{TOK}
\hat{T}(\kappa=+1)\equiv
\hat{{\cal T}}=-2\alpha \hat{P}_\alpha,\quad \hat{P}_\alpha=
\frac{1}{2}\left(
1-\frac{\alpha\hat{p}^i \sigma^i}{\hat{\omega}}\right),
\end{equation}
\begin{eqnarray}
\label{LG1}
\hat{ J}^{ik}(\kappa=+1)\equiv
\hat{{\cal J}}^{ik}&=&-\hat{q}^i \hat{p}^k + \hat{q}^k \hat{p}^i -
\frac{i\hbar}{4}\left[\sigma^i, \sigma^k\right]_{-},\nonumber\\
\hat{ J}^{0k}(\kappa=+1)\equiv\hat{{\cal J}}^{0k}&=&-x^0 \hat{p}^k
-\frac{1}{2}\left[\hat{q}^k, \hat{\omega}\right]_{+}
-\frac{i\hbar}{4 \hat{\omega}}\hat{p}^j \left[\sigma^k.
\sigma^j\right]_{-}.
\end{eqnarray}
where we introduced the projectors $\hat{P}_\alpha=
 \left(\hat{P}_+,\hat{P}_-\right)$.

 As it was pointed above, the operator $\hat{T}$  commutes with
generators $\hat{J}^{\mu\nu}$  for all $D$--s, and hence the
operator $\hat{P}_\alpha$ commutes with generators $\hat{{\cal
J}}^{\mu\nu}$. From this follows, that $\hat{{\cal J}}^{\mu\nu}$
can be represented in the form
\begin{equation}
\label{LOP}
\hat{{\cal J}}^{\mu\nu}=\hat{P}_+
\hat{{\cal J}}^{\mu\nu}+\hat{P}_-\hat{{\cal
J}^{\mu\nu}}=\hat{P}_+\hat{{\cal J}}^{\mu\nu}\hat{P}_+
+\hat{P}_-\hat{{\cal
J}^{\mu\nu}}\hat{P}_-.
\end{equation}
This again is a reflection of the fact, that the corresponding
representation of the Lorentz group is reducible.

Now the condition (\ref{TO2}) for the definite value of the
$\alpha$ takes the form
\begin{equation}
\label{PO}
\hat{P}_\alpha f^1=0.
\end{equation}

Since,  as  it is known, $\hat{p}^i \sigma^i/\hat{\omega}$
is  operator of chirality, the condition (\ref{PO}) extracts
from  $f^1$  the  state with one value of chirality equal to
$\alpha$.

Thus we have constructed in any even--dimensional space--time the
quantum mechanics of the system, describing a neutral particle
(the antiparticle coincides with the particle) which has one
value of chirality (equal to $\alpha$). It is natural to call
it Mayorana--Weyl particle.

Let  us turn now to the clarification of the relation of the
description of the Mayorana--Weyl  particle constructed here to
the Mayorana--Weyl particle in the Dirac picture, which exists
only in the dimensions $D=2({\rm mod}\,8)$ \cite{JS}). We will
consider, for definiteness, the space--time dimension $D=10$.

Let   us  write  the  explicit  representation  of  $\gamma$
--matrices for $D=10$
\begin{equation}
\label{GM}
\Gamma^0=\left(
\begin{array}{cc}
I&0\\0&-I
\end{array}
\right), \quad
\Gamma^i=\left(
\begin{array}{cc}
0&\sigma^i\\-\sigma^i&0
\end{array}
\right), \quad
\Gamma_{11}=\left(
\begin{array}{cc}
0&I\\I&0
\end{array}
\right),
\end{equation}
where  the  matrices  $\sigma^i$ are given by the expressions
\begin{eqnarray}
\label{SM}
&&\sigma^1=i\gamma^2\otimes \gamma^5,
\quad\sigma^2=i\gamma^1\otimes I_4, \quad
\sigma^3=\gamma^5\otimes I_4,
\quad\sigma^4=i\gamma^3\otimes I_4, \nonumber\\
&&\sigma^5=i\gamma^2\otimes \gamma^0, \,
\sigma^6=\gamma^2\otimes \gamma^1, \,
\sigma^7=\gamma^2\otimes \gamma^2, \,
\sigma^8=\gamma^2\otimes \gamma^3, \,
\sigma^9=\gamma^0\otimes I_4.
\end{eqnarray}
Here  $\gamma^5,  \,  \gamma^\mu  \,  (\mu=0, 1, 2, 3)$ are
usual  Dirac  $\gamma$-- matrices, $I_4$ is the unit $4\otimes
4$ matrix. The $\sigma^i$ matrices have the properties, which
can be easily deduced from the representation (\ref{SM}):
\begin{equation}
\label{PSM}
{\sigma^{i}}^T=(-1)^{i+1}\sigma^i, \quad\,
{\sigma^i}^*=(-1)^{i+1}\sigma^i, \,\quad
{\sigma^i}^+=\sigma^i.
\end{equation}
It's  convenient  to  represent $\Gamma$ matrices as a direct
product of a matrices of lower dimensions:
\begin{equation}
\label{DPG}
\Gamma^0=\tau^3\otimes I_4 \otimes I_4, \quad
\Gamma^i=i\tau^2\otimes \sigma^i, \quad i=1, \dots, 9,\quad
\Gamma_{11}=\tau^1\otimes I_4 \otimes I_4
\end{equation}
where $\tau^1,\,\tau^2,\,\tau^3$ are Pauli matrices.

The properties of $\Gamma$ matrices
\begin{equation}
\label{PG}
{\left(\Gamma^0\right)}^{-1}=\Gamma^0, \quad
\left(\Gamma^i\right)^{-1}=-\Gamma^i,
\quad\left(\Gamma_{11}\right)^{-1}=\Gamma_{11},\quad
{\left(\Gamma^i\right)}^{T}=(-1)^i \Gamma^i
\end{equation}
can  be  easily  deduced from (\ref{DPG}) using
(\ref{PSM}).
Write down the expression for the matrix $M\equiv C\Gamma^0$
($C$ is the charge conjugation matrix),
 which is the same in both Dirac and Foldy--Wouthuysen
representations:
\begin{equation}
\label{MM}
M=\Gamma^2\Gamma^4\Gamma^6\Gamma^8\Gamma_{11}=\tau^1\otimes
\Lambda,\quad \Lambda=I_2\otimes\tau^2\otimes I_2\otimes\tau^2.
\end{equation}

Consider now the equation (\ref{PO})
\begin{equation}
\label{WC2}
\left(\frac{\hat{p}^i \sigma^i}{\hat{\omega}}-
\alpha\right)f^1=0
\end{equation}
the  solution  of  which is a Mayorana--Weyl spinor in $D=10$
dimensions  in quantum mechanics. Let us
write the analogues of equations (\ref{WC2}) for the complex
conjugated spinor $f^{1*}$. Taking into account the relation
$\hat{p}_\mu^*=-\hat{p}_\mu\quad(\hat{p}_\mu^+=\hat{p}_\mu)$
we have
\begin{equation}
\label{WEC}
\left(\frac{\hat{p}^i \sigma^{i*}}{\hat{\omega}}
-\alpha\right)f^{1*}=0
\end{equation}

Using the relation
\begin{equation}
\label{LS}
\Lambda{\sigma^i}^*=\sigma^i\Lambda,
\end{equation}
 which can
be easily checked by direct calculations and commuting $\Lambda$
with the operators acting on $f^1$ in equation (\ref{WEC}) we
find
\begin{equation}
\label{WEC1}
\left(\frac{\hat{p}^i \sigma^i}{\hat{\omega}}
+\alpha\right)(\Lambda f^{1*})=0
\end{equation}
The equations (\ref{WC2}) and (\ref{WEC1}) may be written in the
equivalent form
\begin{equation}
\label{MWE}
\hat{T}f_M=0,
\end{equation}
where $\hat{T}$ is given by (\ref{TO}) and the notation
\begin{equation}
\label{MW}
f_M =
 \left(
\begin{array}{c}
f^1\\ \Lambda f^{1*}
\end{array}\right)
\end{equation}
was introduced. It can be proved, that the existence of a real
matrix $\Lambda$ with the property
(\ref{LS})  singles out in QFT the
space--time dimensions $D=2({\rm mod}\,8)$ \cite{JS}.

The $f_M$ defined by (\ref{MW}) is just a Mayorana bispinor in the
Foldy--Wouthuysen representation. In fact,
writing down the condition for  the Mayorana bispinor $f$
\begin{equation}
\label{MC}
f= f^{C}= M f^*
\end{equation}
 we find,  that  the  spinors,  satisfying (\ref{MC}) , have the
form  of (\ref{MW}).

The equation (\ref{MWE}) corresponds to equation (\ref{TO2})
which, as it was found in the previous section, are Weyl
conditions on the Mayorana spinor in the Foldy--Wouthuysen
representation. Furthermore, the Schr\"odinger equation for $f_M$
has the form of (\ref{SE}). The Lorentz generators
for $f_M$ can be constructed through the generators (\ref{LG1})
for  the $f^1$   using the relation
\begin{equation}
\label{MLG}
i\hat{{\cal J}}^{\mu\nu}(\kappa=-1)=-i\Lambda
\hat{{\cal J}}^{\mu\nu*}(\kappa=+1)\Lambda.
\end{equation}

Direct calculations of the right hand side of the (\ref{MLG})
using (\ref{LS}) shows that the \mbox{$\hat{{\cal
J}}^{\mu\nu}(\kappa=-1)$} coincides with $\hat{J}^{\mu\nu}(\kappa=-1)$,
which ensures the Lorentz covariance of Mayorana bispinor
(\ref{MW}).

 Thus using the quantum mechanical column $f^1$ we constructed
 the Mayorana --Weyl bispinor in the Foldy -- Wouthuysen
 representation.

This research was partially supported  by the grant 211-5291 YPI of
the German Bundesministerium f\"ur Forschung und Technologie.
I.V.Tyutin was supported in part by grant \# M21300 from
international Science Foundation and Government of the Russian
Federation and by European Community Commission under contract
INTAS-94-2317.


\end{document}